\documentclass[conference,letter]{IEEEtran}

\usepackage{graphicx} 
\graphicspath{{./figures/}}
\usepackage{float}
\usepackage{array} 
\usepackage{paralist} 
\usepackage{subfig}
\usepackage{fancyhdr} 
\usepackage{amsmath, amsfonts}
\usepackage{mathtools}
\usepackage[most]{tcolorbox}
\usepackage{tikz}
\usepackage{qcircuit}
\newcommand{\ket}[1]{\ensuremath{\left|#1\right\rangle}} 

\begin{document}
    \title{A Simple Quantum Neural Net with a Periodic Activation Function} 
    \author{\IEEEauthorblockN{Ammar~Daskin}
        \IEEEauthorblockA{\textit{Department of Computer Engineering},\\
            \textit{Istanbul Medeniyet University},\\
            Kadikoy, Istanbul, Turkey\\
            Email: adaskin25-at-gmail-dot-com}
    }
    \maketitle
    \begin{abstract}
    In this paper, we propose a simple neural net that requires only $O(nlog_2k)$ number of qubits and $O(nk)$ quantum gates: Here, $n$ is the number of input parameters, and $k$ is the number of weights applied to these parameters in the proposed neural net.
We describe the network in terms of a quantum circuit, and then draw its equivalent classical neural net which involves $O(k^n)$ nodes in the hidden layer. 
Then, we show that the network uses a periodic activation function of cosine values of the linear combinations of the inputs and weights.
    The backpropagation is described through the gradient descent, and then iris and breast cancer datasets are used for the simulations. 
    The numerical results indicate the network can be used in machine learning problems and it may provide exponential speedup over the same structured classical neural net.
 \end{abstract}
    
    \IEEEpeerreviewmaketitle
 
 Neural networks are composed of many non-linear components that mimic the learning mechanism of a human-brain. The training in networks is done by adjusting  weight constants applied to the input parameters.
However, the considered numbers of input parameters and the layers in these networks increase the computational cost dramatically. 
 Quantum computers are believed to be more powerful computational machines which may allow to solve many intractable problems in science and engineering. 
Although building useful quantum computers with many qubits are the main focus of recent experimental research efforts \cite{mohseni2017commercialize}, the complete use of these computers are only possible by novel algorithms that provides  computational speed-up over classical algorithms.

Although many early efforts to describe quantum perceptron (e.g.  \cite{lewenstein1994quantum}) and neural network models (e.g. \cite{altaisky2001quantum,NARAYANAN2000231,behrman2000simulations}) and general discussions on quantum learning \cite{kak1995quantum,chrisley1997learning},  research in quantum machine learning \cite{schuld2014quest,wittek2014quantum,biamonte2017quantum} and quantum big data analysis (e.g. \cite{lloyd2014quantum,rebentrost2014quantum}) gained momentum in recent years. 
Various quantum learning algorithms and subroutines  are proposed(see the review articles \cite{schuld2014quest,wittek2014quantum,biamonte2017quantum} and the survey \cite{arunachalam2017guest} on general quantum learning theory): While many of the recent algorithms are based on variational quantum circuits\cite{peruzzo2014variational,schuld2018circuit,grant2018hierarchical,
mitarai2018quantum,xia2018quantum}, some of them employs quantum algorithms: For instance, Ref.\cite{ricks2004training} uses Grover search algorithm \cite{grover1997quantum} to extract solution from the state which is prepared by directly mapping weights and inputs to the qubits.  
The measurement in the output of a layer is used to decide the inputs to  hidden layers. 
In addition,
Ref.\cite{schuld2015simulating} has used the phase estimation to imitate the output of a classical perceptron where the binary input is mapped to the second register of the algorithm and the weights are implemented by phase gates.   
 The main problem in current quantum learning algorithms is to tap the full power of artificial neural networks into the quantum realm by providing robust data mapping algorithms from the classical realm to the quantum and processing this data in a nonlinear way similar to the classical neural networks.
It is shown that a repeat until success circuit can be used to create a quantum perceptron with nonlinear behavior as a main building block of quantum neural nets \cite{cao2017quantum}. It is also explained in Ref.\cite{schuld2018quantum} how mapping data into Hilbert space can help for kernel based learning algorithms. 

The superposition is one of the physical phenomena that allows us to design computationally more efficient quantum algorithms.  
In this paper, we present a quantum neural net by fully utilizing the superposition phenomenon. 
After describing the network as a quantum circuit, we analyze the quantum state of the circuit-output and show that it relates to a neural net with a periodic activation function involving the cosine values of the weighted sum of the input parameters.
We then present the complexity of the network and then show the numerical simulations for two different data sets.

   \section{Quantum Neural Net}
 In classical neural networks,  linear combinations of  input parameters with different weights are fed into multiple neurons. 
 The output of each neuron is determined by an activation function such as the following one (see Ref.\cite{Nielsen2015} for a smooth introduction): 
  \begin{align}
 \mbox{output} & = & \left\{ \begin{array}{ll}
 0 & \mbox{if } \sum_j w_j x_j \leq \mbox{ threshold} \\
 1 & \mbox{if } \sum_j w_j x_j > \mbox{ threshold}
 \end{array} \right.
\end{align}
Nonlinear activation functions such as hyperbolic and sigmoid functions are more commonly used to make the the output of a neuron smoother: i.e. a small change in any weight causes a small change in the output.
It has been also argued that periodic activation functions may improve the general performance of neural nets in certain applications \cite{Sopena1999,Masahiro1994,MORITA19961477}. 

\begin{figure}[ht]
    \centering 
\mbox{
\Qcircuit @C=1em @R=.7em {
   \lstick{\ket{0}} & \gate{H} & \ctrl{1} & \gate{H} & \qw & \meter& \qw &z\\
   \lstick{\ket{\psi}} & {/} \qw & \gate{  \mathcal{ U}(\omega,x)} & \qw & \qw&\qw{/}& \qw
}
}
\caption{\label{FigCircuit}The proposed quantum neural network with 1-output and $n$-input parameters.}
\end{figure}
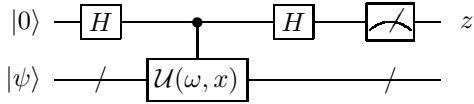

Here, let us first assume that an input parameter $x_j$ is expected to be seen with $k$ number of different weights $\{w_{j1}, \dots, w_{jk}\}$ in the network. 
For each input, we will construct the following operator to represent the input behavior of a parameter $x_j$:
 \begin{equation}
 U_{x_j} = \left(\begin{matrix}
         e^{iw_{j1}x_j} & & & \\
         &  e^{iw_{j2}x_j} & & \\
         &  &\ddots & \\
         & &  &   e^{iw_{jk}x_j} \\
     \end{matrix}\right)
    \end{equation}
Since $U_{x_j}$ is a $k$ dimensional matrix, for each input ${x_j}$, we employ $\log_2k$ number of qubits. Therefore, $n$-input parameters lead to $n$ number of $U_{x_j}$ and require ${n\log_2k}$ number of qubits in total: This is depicted by the following circuit: 
\begin{center}
    \mbox{
        \Qcircuit @C=1em @R=.7em {
            & {/} \qw & \gate{U_{x_1}} & \qw {/} & \qw&\qw& \\
            & {/} \qw & \gate{U_{x_2}} & \qw {/} & \qw&\qw& \\
            && \vdots\\
            & {/} \qw & \gate{U_{x_n}} & \qw {/} & \qw&\qw& \\
        }
    }
\end{center}
We can also describe the above circuit by the following tensor product:
\begin{equation}
   \mathcal{ U}(\omega, x) = U_{x_2} \otimes U_{x_2}\otimes \dots\otimes U_{x_n}.
  \end{equation}
In matrix form, this is equal to:
  \begin{equation}
  \left(\begin{matrix}
    e^{{i\sum_{j}^n w_{j1}x_j}}& & & \\
    & e^{i\sum_{j}w_{j1}x_j + w_{n2}x_n}& & \\
    &  &\ddots & \\
    & &  &   e^{i\sum_{j}w_{jk}x_j} \\
    \end{matrix}\right).
    \end{equation}
 The diagonal elements of the above matrix describe an input with different weight-parameter combinations.  Here, each combination is able to describe a path (or a neuron in the hidden layer) we may have in a neural net. 
 The proposed network with 1-output and n-inputs is constructed by plugging this matrix into the circuit drawn in Fig.\ref{FigCircuit}. 
  
 In the circuit, initializing \ket{\psi} as an equal superposition state allows the system qubits to  equally impact the first qubit which yields the output.
 In order to understand how this might work as a neural net, we will go through the circuit step by step:
  At the beginning, the initial input to the circuit is defined by:
\begin{equation}
    \ket{0}\ket{\psi} = \frac{1}{\sqrt{N}}\ket{0}\sum_j^N\ket{\bf j},
\end{equation}
where $N = k^n$ describing the matrix dimension and \ket{\bf j} is the $j$th vector in the standard basis.
After applying the Hadamard gate and the controlled $U(\omega,x)$ to the first qubit, the state becomes
\begin{equation}
 \frac{1}{\sqrt{2N}}\left(\ket{0}\sum_i^N\ket{\bf j}+\ket{1}\sum_j^Ne^{i\alpha_j}\ket{\bf j}\right).
\end{equation}
Here, $\alpha_j$ describes the phase value of the $j$th eigenvalue of $\mathcal{U}$.
After the second Hadamard gate, the final state reads the following:
\begin{equation}
\frac{1}{2\sqrt{N}}\left(\ket{0} \sum_j^N\left(1+ e^{i\alpha_j}\right)\ket{\bf j}
+\ket{1} \sum_j^N\left(1- e^{i\alpha_j}\right)\ket{\bf j}\right).
\end{equation}
If we measure the first qubit, the probability of seeing \ket{0} and \ket{1}, respectively $P_0$ and $P_1$, can be obtained from the above equation as:

\begin{align}
P_0 = \frac{1}{4N}\sum_j|1+e^{i\alpha_j}|^2=\frac{1}{2N}\sum_j^N\left(1+cos(\alpha_j)\right),\\
 P_1 = \frac{1}{4N}\sum_j|1-e^{i\alpha_j}|^2=\frac{1}{2N}\sum_j^N\left(1-cos(\alpha_j)\right).\\
\end{align}
If a threshold function is applied to the output, then 
  \begin{align}
z & = \left\{ \begin{array}{ll}
  0 & \mbox{if } P_1 \leq  P_0 \\
  1 & \mbox{if } P_1 > P_0
  \end{array} \right.
  \end{align}
Here, applying the measurement a few times, we can also obtain enough statistics for $P_0$ and $P_1$; and therefore describe $z$ as the success probability of the desired output: i.e., $z= P_d$.

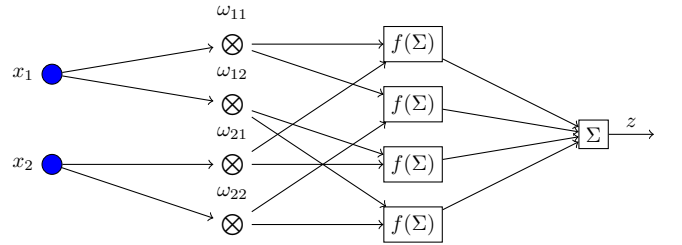
\begin{figure}[h!]
    \centering 
    \begin{tikzpicture}
    [ scale=0.8,every node/.style={scale=0.8},  cnode/.style={draw=black,fill=#1,minimum width=2mm,circle},
    ]
    \tikzstyle{sqrG}=[draw=black,fill=white,minimum width=3mm,rectangle];
    \tikzstyle{sqrR}=[draw=black,fill=red!25,minimum width=3mm,rectangle];
    \node[sqrG] (s11) at (6,-1) {$f(\Sigma)$};
    \node[sqrG] (s12) at (6,-2) {$f(\Sigma)$};
    \node[sqrG] (s21) at (6,-3) {$f(\Sigma)$};
    \node[sqrG] (s22) at (6,-4) {$f(\Sigma)$};
    \node[sqrG,fill=white] (out) at (9,-2.5) {$\Sigma$};
    \draw [->](out) -- ++(1,0) node[above, midway] {$z$};
    \foreach \x in {1,...,2}
    {   \pgfmathparse{\x<4 ? \x : "n"}
        \node[cnode=blue,label=180:$x_{\pgfmathresult}$] (x-\x) at (0,{-1.5*\x-div(\x,4)}) {};
        \node[label=90:$\omega_{1\pgfmathresult}$] (p-1\x) at (3,{-\x-div(\x,4)}) {$ \bigotimes$};
        \node[label=90:$\omega_{2\pgfmathresult}$] (p-2\x) at (3,{-2-\x-div(\x,8)}) {$ \bigotimes$};
    }
    \foreach \x in {1,...,2}
    {   \foreach \y in {1,...,2}
        {   
            \draw[->] (p-1\x) -- (s\x\y);
            \draw[->] (p-2\y) -- (s\x\y);
            \draw[->] (s\x\y)--(out);
        }
    }
    \foreach \x in {1,...,2}
    {   \foreach \y in {1,...,2}
        {   \draw[->] (x-\x) -- (p-\x\y);
        }
    }
    \end{tikzpicture}
    \caption{\label{Fignet}The equivalent representation of the quantum neural net for two input parameters and two weights for each input: i.e. $n=2$ and $k=2$.}
\end{figure}

The whole circuit can be also  represented as an equivalent neural net shown in Fig.\ref{Fignet}. In the figure, $f$ is the activation function described by:
 \begin{equation}
     f(\alpha) = 1- cos(\alpha).
    \end{equation}

\subsection{The Cost Function}
We will use the following to describe the cost of the network for one sample:
\begin{equation}
    C = \frac{1}{2s}\sum_j^s(d_j-z_j)^2,
\end{equation}
where $d_j$ is the desired output for the $j$th sample and $s$ is the size of the training dataset.
\subsection{Backpropagation with Gradient Descent}
The update rule for the weights  is described by the following:
\begin{equation}
    \omega_i = \omega_i-\eta \frac{\partial C}{\partial w_i}.
\end{equation}
Here, the partial derivative can be found via chain rule: 
For instance, from Fig.\ref{Fignet} with an input $\{x_1, x_2\}$, we can obtain the gradient for the weight $\omega_{11}$  as (the constant coefficients omitted): 
\begin{equation}
\label{EqGradient}
  \frac{\partial C_j}{\partial \omega_{11}} = 
  \frac{\partial C_j}{\partial z_j} 
  \frac{\partial z_j}{\partial \alpha}
  \frac{\partial  \alpha}{\partial \omega_{11}} \approx  (d_j-z_j) P_{d_j}^2x_1.
\end{equation}
\section{Complexity Analysis}
The computational complexity of any quantum algorithm is determined by the number of necessary single and CNOT gates and the number of qubits.
The proposed network in Fig.\ref{FigCircuit} only uses $nlog_2k+1$ number of qubits.
In addition, it only has $nk$ controlled phase gates ($k$ number of gates for each input.) and two Hadamard gates.
Therefore, the  complexity is bounded by $O(nk)$.

A simulation of the same network on classical computers would require exponential overhead since the size of $\mathcal{U}(\omega,x)$ is $k^n$ and the classical equivalent network involves $k^n$ neurons in the hidden layer.
 Therefore, the proposed quantum model may provide exponential speed-up for certain structured networks.
 
\section{Simulation of the Networks for Pattern Recognition}
The circuit given Fig.\ref{FigCircuit} is run for two different simple data sets: breast cancer(699 samples)  and iris flowers(100 samples for two flowers) datasets (see \cite{Dua:2017} for datasets).  For iris-dataset we only use the samples for two flowers. 
The input parameters are mapped into the range $[-1,0]$. Then, for each $\eta$ value and dataset, 80\% of the whole sample dataset is randomly chosen for training, the remaining 20\% of the dataset is used for testing. 

\begin{figure*}
\includegraphics[width=7in]{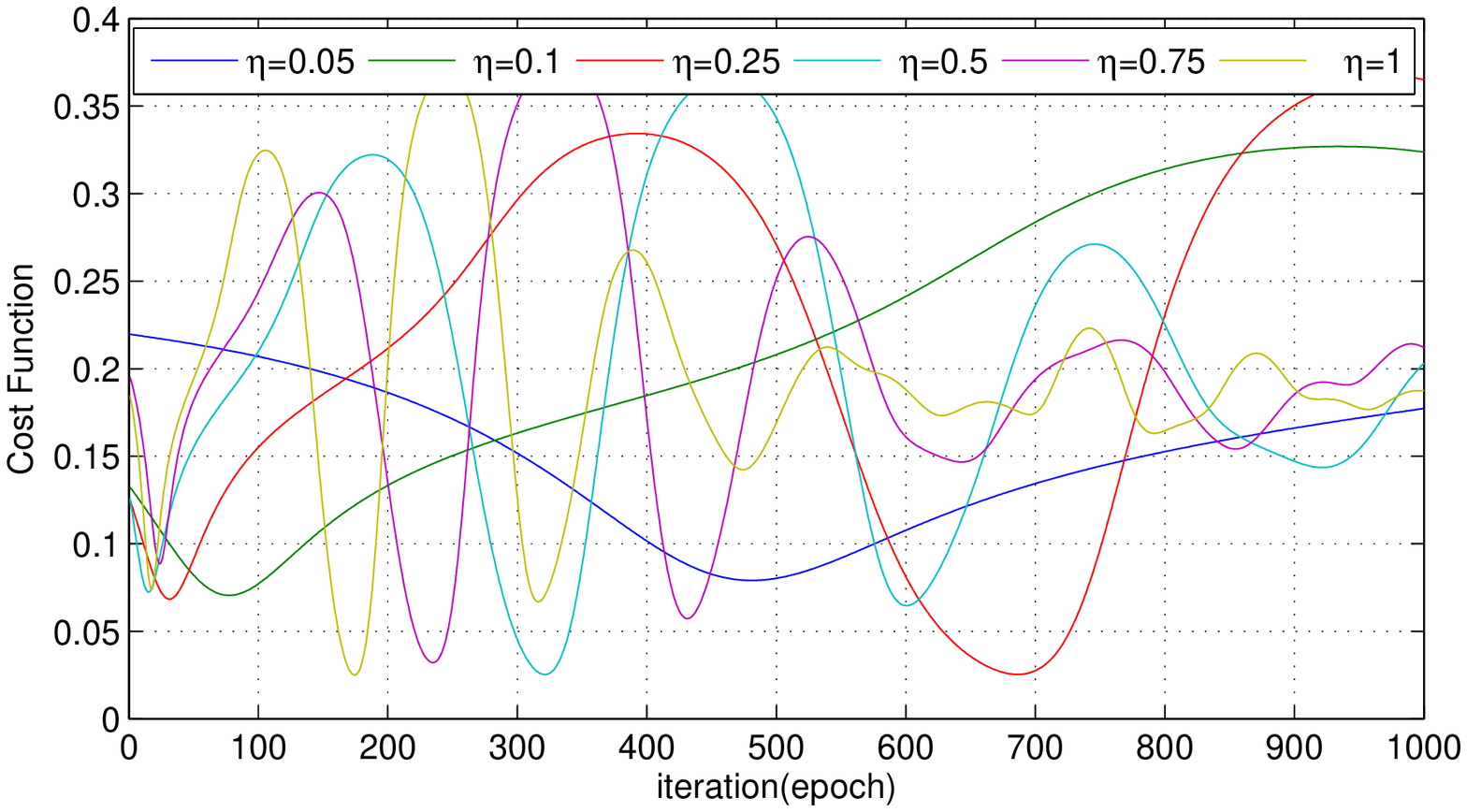}
\caption{Evaluations of the cost function with different learning rates for iris flowers dataset.  \label{FigIrisCF}}
\end{figure*}

\begin{figure*}
\includegraphics[width=7in]{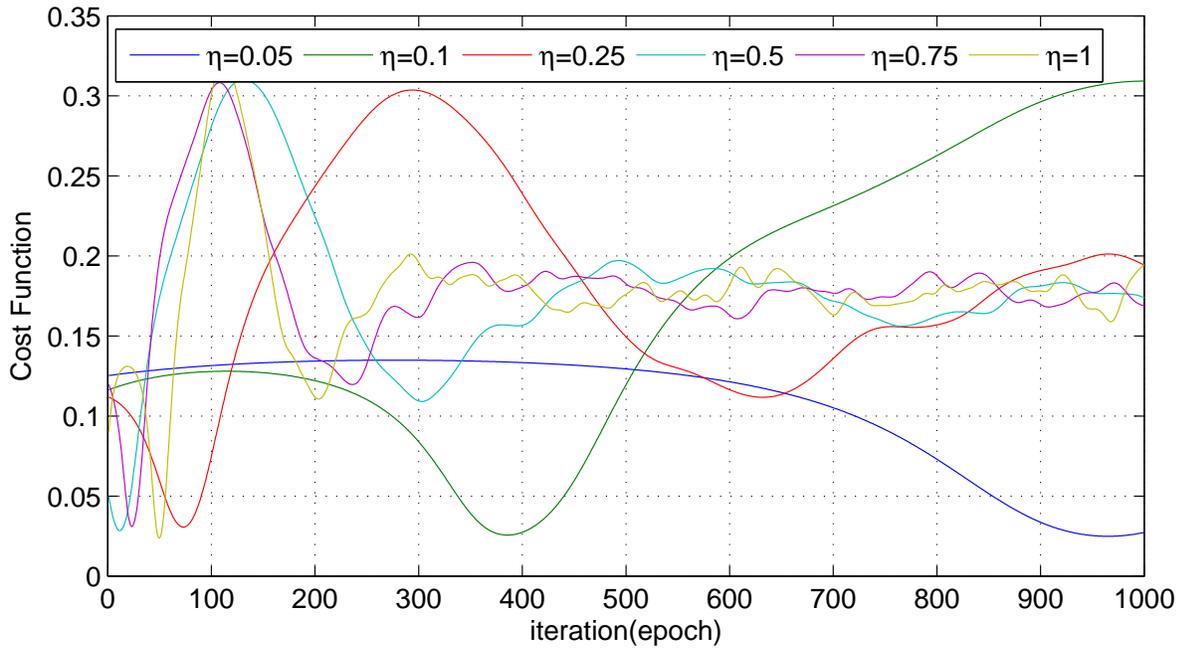}
\caption{Evaluations of the cost function with different learning rates for breast cancer dataset.  \label{FigCancerCF}}
\end{figure*}

Fig.\ref{FigIrisCF} and Fig.\ref{FigCancerCF} show the evaluations of the cost function in each epoch (batch learning is used).
Since $(d_j-z_j)$ in Eq.\eqref{EqGradient} is always positive and because of the periodicity of the activation function, as expected the cost function oscillates between maximum and minimum points and finally settle (if the iteration number is large enough) at some middle point.
 
The accuracy of each trained network is also listed in TABLE I. As seen from the table, the network is able to almost completely differentiate the inputs belong to two different classes.
\begin{table}[h]
\caption{Accuracy of the network trained with different learning rates}
\begin{tabular}{l|llllll}
$\eta$& 0.05& 0.1& 0.25& 0.5& 0.75& 1\\
\hline \\
Iris(test)&99\%&   \textbf{100\%}&   99\%&    99\%&   \textbf{100\%}&    99\%\\
Iris(whole) &91\%&   \textbf{98\%}&   95\%&    95\%&   95\%&    95\%\\
Cancer(test)& 97.8\%&   \textbf{98.9\%}&   96.4\%&   97.1\%&   95\%&   96.4\%\\
Cancer(whole)&95.4\%&   96.7\%&   \textbf{96.9\%}&   95.9\%&   95.3\%&   96.1\%\\ 
\end{tabular}
\end{table}

\section{Discussion}

\subsection{Adding Biases}
Biases can be added to a few different places in Fig.\ref{FigCircuit}. 
As an example, for input $x_j$, we can apply a gate $U_{b_j}$ with diagonal phases representing biases to $U_{x_j}$. One can also add a bias gate to the output qubit before the measurement. 
 \subsection{Generalization to Multiple Output}
\begin{figure}[ht]
    \centering 
\mbox{
\Qcircuit @C=1em @R=.7em {
   \lstick{\ket{0}} & \gate{H} & \ctrl{4} & \qw &\qw &\qw &\qw& \gate{H} & \qw & \meter& \qw &z_1\\
   \lstick{\ket{0}} & \gate{H} & \qw & \ctrl{3}&\qw &\qw&\qw & \gate{H} & \qw & \meter& \qw &z_2\\
   \lstick{\vdots} & \vdots &  &  &  &  &&\vdots & &\vdots&\rstick{\vdots} \\
     \lstick{\ket{0}} & \gate{H} & \qw &\qw &\qw &\qw &\ctrl{1} & \gate{H} & \qw & \meter& \qw &z_m\\
   \lstick{\ket{\psi}} & {/} \qw & \gate{  \mathcal{ U}_1} &\gate{  \mathcal{ U}_2}&\dots& & \gate{\mathcal{ U}_m} & \qw & \qw &\qw{/}& \qw
}
}
\caption{\label{FigCircuitGeneralized}The generalized neural net with $m$-output and $n$-input parameters.}
\end{figure}
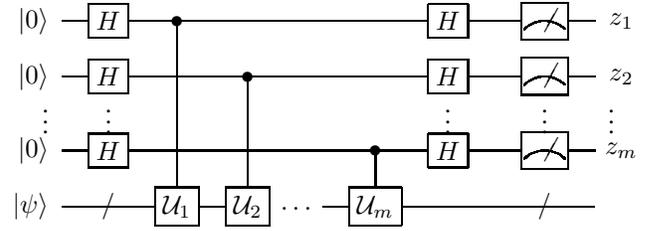

 Different means may be considered to generalize the network for multiple outputs.
As shown in Fig.\ref{FigCircuitGeneralized}, one can generalize the network by sequential applications of $\mathcal{U}_j$s. Here, a $\mathcal{U}_j$ represents a generalized multi-qubit phase gate controlled by the $j$th qubit representing the $j$th output. 
In the application of phase gates, the phases are kicked back to the control qubit. 
Therefore, although   all $\mathcal{U}_j$s operate on the same qubit, the gradient for the parameters of each $\mathcal{U}_j$ is independent since the phases are kicked back to the different control qubits. 
 
\section{Conclusion}
In this paper, we have presented a quantum circuit which can be used to efficiently represent certain structured neural networks. 
While the circuit involves only $O(nk)$ number of quantum gates, the numerical results show that it can be used in machine learning problems successfully. 
We showed that since the simulation of the equivalent classical neural net involves $k^n$ neurons, the presented quantum neural net may provide exponential speed-up in the simulation of certain neural network models.
\bibliographystyle{IEEEtran}
\bibliography{refs}

\end{document}